\documentclass[prd,twocolumn,floatfix,showpacs]{revtex4}
\usepackage{graphicx}
\usepackage{bm}

 \newcommand{\nc}{\newcommand}
 \nc{\mb}[1]{\makebox[#1]{}}
 \nc{\V}{{\rm v}}
 \nc{\scV}{{\scriptscriptstyle V}}
 \nc{\W}{{\scriptscriptstyle W}}
 \nc{\X}{{\scriptscriptstyle X}}
 \nc{\Z}{{\scriptscriptstyle Z}}
 \nc{\D}{{\scriptscriptstyle D}}
 \nc{\F}{{\scriptscriptstyle F}}
 \nc{\G}{{\scriptscriptstyle G}}
 \nc{\A}{{\scriptscriptstyle A}}
 \nc{\N}{{\scriptscriptstyle N}}
 \nc{\Ss}{{\scriptscriptstyle S}}
 \nc{\CSV}{{\scriptscriptstyle CSV}}
 \nc{\CS}{{\scriptscriptstyle CS}}
 \nc{\SV}{{\scriptscriptstyle SV}}
 \nc{\PV}{{\scriptscriptstyle PV}}
 \nc{\be}{\begin{equation}}
 \nc{\ee}{\end{equation}}
 \nc{\bea}{\begin{eqnarray}}
 \nc{\eea}{\end{eqnarray}}
 \nc{\ra}{{\rightarrow}}
 \nc{\ppg}{\pi^+\pi^-\gamma}
 \nc{\Rcsv}{{R_{\CSV}}}

 \def\Gf{{G_{\F}}}
 \def\gFsq{{\Gf^2}}

 \nc{\ovnu}{{\overline{\nu}}}
 \nc{\nuN}{{\nu N_0}}
 \nc{\nubN}{{\ovnu N_0}}
 \nc{\delmn}{{\delta_{m_{\N}}}}
 \nc{\ovdlmn}{{\overline{\delta}_{m_{\N}}}}
 \nc{\snuNC}{{\langle \sigma^{\nuN}_{\NC}\rangle }}
 \nc{\snubNC}{{\langle \sigma^{\nubN}_{\NC}\rangle }}
 \nc{\snuCC}{{\langle \sigma^{\nuN}_{\CC}\rangle }}
 \nc{\snubCC}{{\langle \sigma^{\nubN}_{\CC}\rangle }}
 \nc{\Rnu}{{R^{\nu}}}
 \nc{\Rnub}{{R^{\ovnu}}}
 \nc{\Ftp}{{F_2^{\mu p}}}
 \nc{\Ftn}{{F_2^{\mu n}}}
 \nc{\sintW}{{\sin^2 \theta_{\W} }}
 \nc{\costW}{{\cos \theta_{\W} }}
 \nc{\costtW}{{\cos^2 \theta_{\W} }}
 \nc{\WTdelt}{{\widetilde{\delta}}}
 \nc{\mWsq}{{M_{\W}^2}}
 \nc{\mZsq}{{M_{\Z}^2}}
 \nc{\xF}{{x_{\F}}}
 \nc{\sigS}{{\sigma_{\Ss}}}
 \nc{\Afb}{{A_{fb}}}
 \nc{\vp}{{\bf p}}
 \nc{\rz}{{1\over \rho_0^2}}
 \nc{\uv}{{u_{\V}}}
 \nc{\dv}{{d_{\V}}}
 \nc{\sv}{{s_{\V}}}
 \nc{\cv}{{c_{\V}}}
 \nc{\Uv}{{U_{\V}}}
 \nc{\Dv}{{D_{\V}}}
 \nc{\Sv}{{S_{\V}}}
 \nc{\yW}{{y_{\W}}}
 \nc{\NoA}{{\frac{N}{A}}}
 \nc{\ZoA}{{\frac{Z}{A}}}
 \nc{\NotA}{{\frac{N}{2A}}}
 \nc{\ZotA}{{\frac{Z}{2A}}}
 \nc{\nuf}{{\nu_f}}
 \nc{\delu}{{\delta u}}
 \nc{\deld}{{\delta d}}
 \nc{\dels}{{\delta s}}
 \nc{\dwtilm}{{\delta \widetilde{m}}}
 \nc{\delCSV}{{\delta^{(\CSV)}}}
 \nc{\qbar}{{\overline{q}}}
 \nc{\ubar}{{\overline{u}}}
 \nc{\dbar}{{\overline{d}}}
 \nc{\sbar}{{\overline{s}}}
 \nc{\dubar}{{\delta\ubar}}
 \nc{\ddbar}{{\delta\dbar}}
 \nc{\dsbar}{{\delta\sbar}}
 \nc{\guv}{{g_{\scV}^u}}
 \nc{\gdv}{{g_{\scV}^d}}
 \nc{\gua}{{g_{\A}^u}}
 \nc{\gda}{{g_{\A}^d}}
 \nc{\pis}{{\pi_{\Ss}}}
 \nc{\piv}{{\pi_{\V}}}
 \nc{\piz}{{\pi^0}}
 \nc{\xpi}{{x_\pi}}
 \nc{\deluv}{{\delta \uv}}
 \nc{\deldv}{{\delta \dv}}
 \nc{\delsv}{{\delta \sv}}
 \nc{\als}{{\alpha_{\Ss}}}
 \def\CC{{\scriptscriptstyle CC}}
 \def\NC{{\scriptscriptstyle NC}}
 \def\CS{{\scriptscriptstyle CS}}

 \def\GLS{{\scriptscriptstyle GLS}}

 \def\MSb{{\overline{MS}}}

 \def\IE{{\it i.e.}}
 
 \def\EA{{\it et al.}}

\graphicspath{%
    {converted_graphics/}
    {/}
}
 
\begin{document}

\preprint{ADP-10-16/T712}

 \title{Additional Corrections to the Gross-Llewellyn Smith Sum Rule} 
 \author{J.T. Londergan} 
\affiliation{ Dept. of Physics and Center for Exploration of Energy and 
  Matter, Indiana University, Bloomington, IN 47405}
 
 \author{A.W. Thomas}
\affiliation{ 
 CSSM, School of Chemistry and Physics, University of 
  Adelaide, Adelaide SA 5005, Australia} 
  \date{\today}

 \begin{abstract}
We investigate some QCD corrections that contribute to the Gross--Llewellyn 
Smith (GLS) sum rule, but have not been included in previous analyses 
of it. We first review the techniques by which  
the $xF_3$ structure function is extracted from combinations of 
neutrino and antineutrino cross sections. Next we investigate 
corrections to the GLS sum rule, with particular attention to contributions 
arising from strange quark distributions  
and from charge symmetry violating (CSV) parton distributions. We find that 
additional corrections from strange quarks and parton CSV are likely to 
have a small but potentially significant role in decreasing the current 
discrepancy between the experimental and theoretical estimates of the Gross 
Llewellyn Smith sum rule.
\end{abstract}  

\maketitle

\section{Introduction\label{Sec:Intro}}

The Gross Llewellyn Smith (GLS) sum rule (sometimes referred to as the 
\textit{baryon sum rule}) is obtained from an integral of 
the $xF_3$ structure function obtained in charged-current deep inelastic 
scattering (DIS) from neutrinos and antineutrinos on nucleon or nuclear 
targets \cite{GLS,Hin96}. In recent years the most precise neutrino data 
has been obtained by the CCFR and NuTeV collaborations, from interactions 
of neutrinos and antineutrinos with an iron target \cite{Con98}. 
From measurements by the CCFR group \cite{CCFR}, values have been obtained 
for the GLS sum rule. At the value $Q^2 = 3$ GeV$^2$, the CCFR analysis 
claimed a precision of roughly 3\%. The CCFR data could also be used to 
obtain the GLS sum rule as a function of $Q^2$; this allows one to test 
contributions from higher-order QCD corrections \cite{larin} 
and from higher-twist terms \cite{braun}. The current status of various DIS 
sum rules has been summarized in a review article by Hinchliffe and 
Kwiatkowski \cite{Hin96}.   

In this paper, we point out that some additional QCD effects, 
particularly contributions from strange quarks and parton charge symmetry 
violation (CSV), have not been included to date in estimates of the 
GLS sum rule. One now has recent experimental data on strange quark 
parton distributions, in particular on the asymmetry between strange and 
antistrange quarks, which is relevant for the GLS sum rule. In addition, 
there is much interest in the possibility of CSV in the parton  
distributions~\cite{Londergan:2009kj,Londergan:1998ai,MRST03,Bentz:2009yy}.   

We examine the possible contributions of these terms to the Gross-Llewellyn 
Smith sum rule, finding that such corrections are likely to produce small 
but potentially significant effects. At present, 
theoretical estimates of the GLS sum rule lie one or two standard deviations 
below the data~\cite{Hin96}. We point out that contributions from strange 
quarks and partonic CSV are likely to improve the agreement between theory 
and experiment. 

Our paper is organized as follows. In Sect.~\ref{Sec:Xsects} we review 
the form of neutrino cross sections and the derivation of the GLS sum 
rule. The experimental results of the CCFR group \cite{CCFR} are summarized 
and compared with the theoretical calculations of Hinchliffe and Kwiatkowski 
\cite{Hin96}. In Sect.~\ref{Sec:Extract} we review how the structure 
functions, particularly $xF_3$, are extracted 
from experimental data. We pay special attention to contributions from 
strange quarks and partonic CSV. In Sect.~\ref{Sec:sCSV}, we make estimates 
of these contributions to the GLS sum rule. 

In Sect.~\ref{Sec:GLSnuc}, we review isoscalar corrections to the 
data, which arise because iron is not an isoscalar target. Because of the 
way in which isoscalar corrections have been implemented in previous analyses, 
it is difficult for us to give a definitive, quantitative estimate of 
the contribution of strange quarks and partonic charge symmetry violation 
in the GLS sum rule. Nevertheless, our analysis clearly establishes that 
these corrections can be as large as the quoted errors on the GLS 
sum rule and that they tend to improve the agreement between theory and 
experiment. Our results show that these corrections should be included in 
future analyses.

\section{Neutrino Cross Sections and the Gross-Llewellyn Smith Sum Rule 
\label{Sec:Xsects}}

\begin{figure}[ht] 
\includegraphics[width=1.9in]{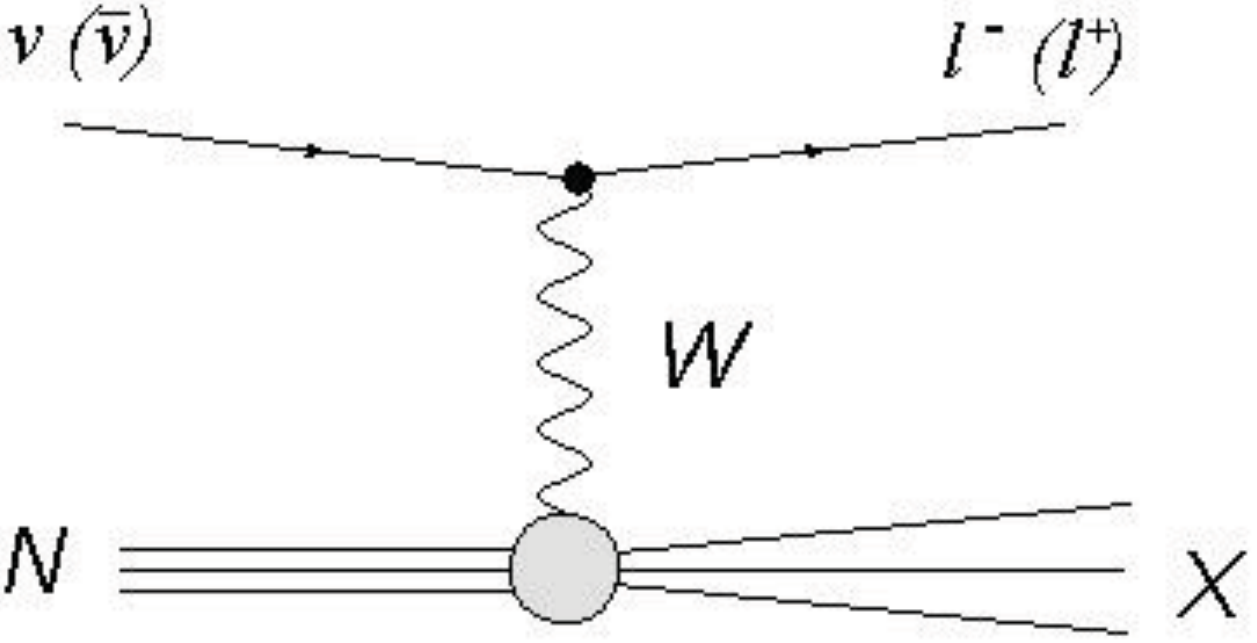}
  \caption{Schematic picture of charged-current amplitudes in DIS induced by 
neutrinos or antineutrinos.}
\label{fig21}
\end{figure}

The cross section for charged current (CC) interactions initiated by 
neutrinos or antineutrinos on nucleons on a proton is shown schematically 
in Fig.~\ref{fig21}. It has the form \cite{Con98,Londergan:2009kj} 
\bea
  \frac{d^2\sigma^{\nu (\bar\nu)\,p}_{\CC}}{dx\,dy} &=& 
  \frac{\gFsq M E_{\nu}}{\pi} \bigl[ f_1(y) F_2^{W^\pm p}(x, Q^2)  
  \nonumber \\ &\pm& f_2(y) xF_3^{W^\pm p}(x, Q^2) \bigr]\ ; 
  \nonumber \\ f_1(y) &=& 1 - y - \frac{xyM^2}{s}+ 
  \frac{y^2}{2}\frac{1+4M^2x^2/Q^2}{1+R_L^{\nu}(x,Q^2)} \nonumber \\ 
  &\approx& 1 - y + \frac{y^2}{2(1+ R_L^{\nu})} \ ; \nonumber \\ 
  f_2(y) &=& y - \frac{y^2}{2} \ .
\label{eq:sig2CC}
\eea  
The relativistic invariants in Eq.~(\ref{eq:sig2CC}) are $Q^2 =-q^2$, the 
square of the four momentum transfer for the reaction, $x$ and $y$.  For four
momentum $k$ ($p$) for the initial state lepton (nucleon), 
we have the relations 
\be 
  x = {Q^2 \over 2p\cdot q}; \qquad y = {p\cdot q \over p\cdot k};   
  \qquad s = (k+p)^2 \ . 
\label{kinem}  
\ee

Eq.~(\ref{eq:sig2CC}) applies in the limit $Q^2 << \mWsq$. We have introduced 
the Fermi coupling constant, $\Gf$, in terms of the electromagnetic 
coupling constant $\alpha$, the $W$ boson mass $M_{\W}$, and the weak mixing  
angle $\theta_{\W}$,  
\be 
 \Gf = \frac{\pi \alpha}{\sqrt{2}\sintW \mWsq} \ .
\label{eq:GFermi}
\ee

For neutrino-induced interactions, $F_2^{W^+ p}(x, Q^2)$ represents the 
$F_2$ structure function corresponding to a $W^+$ being absorbed on the 
proton. In Eq.~(\ref{eq:sig2CC}) we have written the structure function 
$F_1$ in terms of the structure function $F_2$ and the longitudinal to 
transverse cross section ratio $R_L^{\nu}$, 
\IE 
 \be
  2x\,F_1^{W^+ p}(x,Q^2) = \frac{1+4M^2x^2/Q^2}{1+R^{\nu}_L(x,Q^2)} 
  F_2^{W^+ p}(x,Q^2) .
 \label{Rdenn}
 \ee

For the CC reactions initiated by neutrinos, the 
experimental values for $R^{\nu}_L$ are summarized in the review article 
by Conrad, Shaevitz and Bolton \cite{Con98}. 

The most accurate neutrino and antineutrino cross sections are on nuclear 
targets, particularly (in the case of the CCFR and NuTeV measurements) on 
iron targets. The structure functions are typically described in terms of 
nuclear parton distributions. At sufficiently high values of $Q^2$, the 
structure functions per 
nucleon for a nucleus with $Z$ protons and $N = A-Z$ neutrons can be written 
in terms of averages and differences of structure functions for neutrinos 
and antineutrinos. 

Therefore we define 
\bea
\overline{F}_2^{W A}(x) &=& \ZotA \left(F_2^{W^+ p}(x) + F_2^{W^+ p}(x)\right)
  \nonumber \\ &+& \NotA\left(F_2^{W^+ n}(x) + F_2^{W^- n}(x)\right) \ ; 
  \nonumber \\ 
x\overline{F}_3^{W A}(x) &=& \ZotA \left( xF_3^{W^+ p}(x) 
   + xF_3^{W^- p}(x)\right)
   \nonumber \\ &+& \NotA \left( xF_3^{W^+ n}(x) + xF_3^{W^- n}(x)\right)\ ; 
  \nonumber \\ 
\Delta F_2^{W A}(x) &=& \ZoA (F_2^{W^+ p}(x) - F_2^{W^- p}(x)) 
   \nonumber \\ &+& \NoA (F_2^{W^+ n}(x) - F_2^{W^- n}(x)) \ ; \nonumber \\ 
\Delta xF_3^{W A}(x) &=& \ZoA (xF_3^{W^+ p}(x) - xF_3^{W^+ p}(x)) 
   \nonumber \\ &+& \NoA (xF_3^{W^+ n}(x) - xF_3^{W^+ n}(x)) \ .
  \nonumber \\  
\label{eq:FAdef}
\eea

In terms of parton distribution functions it is straightforward to show 
that 
\bea  
\overline{F}_2^{W A}(x) &=& x(u^+(x) + d^+(x) + s^+(x) + c^+(x)) \nonumber \\ 
  &-& x\NoA (\delta u^+(x) + \delta d^+(x)) \ ; \nonumber \\ 
x\overline{F}_3^{W A}(x) &=& x(u^-(x) + d^-(x) + s^-(x) + c^-(x)) \nonumber \\ 
  &-& x\NoA (\delta u^-(x) + \delta d^-(x)) \ ; \nonumber \\ 
\Delta F_2^{W A}(x) &=& 2x[s^-(x) - c^-(x) + \nuf (u^-(x) \nonumber \\ 
  &-& d^-(x)) + \NoA (-\delta u^-(x) + \delta d^-(x))] \ ; \nonumber \\ 
\Delta xF_3^{W A}(x) &=& 2x[s^+(x) - c^+(x) + \nuf (u^+(x) \nonumber \\ 
  &-& d^+(x)) + \NoA (-\delta u^+(x) + \delta d^+(x)) ] \ . 
  \nonumber \\ 
\label{eq:FApdf}
\eea
In Eq.~(\ref{eq:FApdf}) we have assumed the impulse approximation, 
\IE~that the nuclear structure functions are simply given as the sum of 
free nucleon parton distributions. Nuclear modifications of the parton 
distributions have been considered by various groups 
\cite{Kum02,Hir04,Hir05,Kul06,Kul07,Kul07b}, and we note 
particularly the recent discovery of an isovector EMC 
effect~\cite{Cloet:2009qs}. In Eq.~(\ref{eq:FApdf}) we use the notation 
 \be
 q^{\pm}(x) \equiv q(x) \pm \bar{q}(x) \ .
\label{eq:qpm}
 \ee 
We have introduced the neutron asymmetry parameter 
$\nuf = (N-Z)/A$. For the CCFR and NuTeV iron targets the average value is 
$\nuf = 0.0567$ \cite{Sel97,Sel97a}. We also include possible parton charge 
symmetry violation (CSV), with the notation 
\bea
 \delta u(x) &=& u^p(x) - d^n(x) ; \nonumber \\ 
 \delta d(x) &=& d^p(x) - u^n(x) ,  
\label{eq:CSVdef}
\eea 
and an analogous equation for antiquarks. 

The neutrino/antineutrino average structure functions contain contributions 
from light quarks including the relatively large light valence quark 
contributions. The structure function differences on a nuclear target  
contain contributions from heavy quarks, partonic CSV contributions, 
and light quark contributions proportional to the neutron asymmetry 
$\nuf$.  In previous work the term 
$\Delta xF_3^{W A}(x)$ has been treated as a small perturbation, while the 
term $\Delta F_2^{W A}(x)$ has 
been neglected. In this paper we will examine the possible 
effects of the term $\Delta F_2^{W A}(x)$ on the Gross-Llewellyn Smith sum 
rule.  
 
\subsection{The Gross-Llewellyn Smith Sum Rule}
 \label{Sec:GLS}

 The Gross-Llewellyn Smith (GLS) Sum Rule \cite{GLS} is derived from the
first moment of the $F_3$ structure functions for neutrinos and 
antineutrinos.  
The easiest derivation of the Gross-Llewellyn Smith sum rule results from 
summing the $xF_3$ structure 
functions for neutrinos and antineutrinos on a proton. Using 
Eq.~(\ref{eq:FApdf}) for the case of a proton ($Z=1, N=0$) we obtain 
\bea
S_{\GLS} &\equiv& \int_0^1 \frac{dx}{x}\,x\overline{F}_3^{W p}(x) = 
\int_0^1 [ u^-(x) + d^-(x) \nonumber \\ &+& s^-(x) + c^-(x)] 
  \, dx = 3
\label{eq:GLSnaive}
\eea
In Eq.~(\ref{eq:GLSnaive}) we have neglected various QCD corrections and  
higher twist contributions. Without these corrections, the GLS sum rule is 
equal to three because the first moment of the light quark parton 
distributions gives the total number of valence quarks in the nucleon. 
From valence quark normalization, the 
first moments of the strange and charm quark asymmetries, $s^-(x)$ and 
$c^-(x)$ respectively, vanish when integrated over all $x$. 

In order to compare the Gross-Llewellyn Smith sum rule with experimental 
data, we must consider a number of corrections. For pedagogical purposes 
we will discuss the GLS sum rule for an isoscalar target. At the end of 
this article we will review the corrections that are made for 
non-isoscalar targets such as iron. After applying a series of QCD   
corrections one obtains     
 \bea
&\,& S_{\GLS}^{iso} \equiv \int_0^1 \frac{dx}{x} \,
  x\overline{F}_3^{W A}(x) 
  \nonumber \\ &=& 3 \, \biggl[ 1 - 
  \frac{\als(Q^2)}{\pi} - a(n_f)\left(\frac{\als(Q^2)}{\pi}\right)^2  
  \nonumber \\ &-& b(n_f)\left(\frac{\als(Q^2)}{\pi}\right)^3 \biggr] 
  + \Delta HT.  
 \label{SGLSdef}
 \eea
 The naive Gross-Llewellyn 
 Smith sum rule is correct only in leading twist approximation, and only to 
 lowest order in the strong coupling constant $\als$. Our expression for 
 the GLS sum rule thus includes a QCD correction (the term in square
 brackets in Eq.~(\ref{SGLSdef})), which was derived by Larin and Vermaseren 
\cite{larin} using a QCD scale parameter $\Lambda_{QCD} = 213 \pm 50$ MeV, and 
the quantity $\Delta HT$
 represents a higher twist contribution \cite{braun}. This is 
summarized in the review article on QCD sum rules by Hinchliffe and 
Kwiatkowski \cite{Hin96}. 

 As is the case for the Adler \cite{Adl66} and Gottfried \cite{Got67} sum 
 rules, the Gross-Llewellyn Smith sum rule requires that the structure
 function be divided by $x$ in performing the integral.  This
 gives a strong weighting to the small-$x$ region, such that
 as much as 90\% of the sum rule comes from the region $x \leq 0.1$.
 The most precise value has
 been obtained by the CCFR collaboration \cite{CCFR}, which
 measured neutrino and antineutrino cross sections on an iron
 target, using the quadrupole triplet beam (QTB) at Fermilab.
 A summary of experimental details for precision measurements using  
 high-energy neutrino beams is given in the review article by Conrad, 
 Shaevitz and Bolton~\cite{Con98}, and a detailed description 
of the experimental details and analysis procedure used by the CCFR 
collaboration is given in the thesis of Seligman~\cite{Sel97a}. 

Because of the large contribution to the GLS 
 sum rule from small $x$, one measures $xF_3$ at various values of $x$, 
 and evaluates the integral 
\be
S_{\GLS}(x) = \int_x^1 \frac{dy}{y} \, y\overline{F}_3^{W A}(y) .
\label{eq:SGLSx} 
\ee
The Gross-Llewellyn Smith sum rule is then obtained by taking the 
limit  
\be
S_{\GLS} = {\rm lim}_{x \rightarrow 0} \ S_{\GLS}(x) \ .
\label{eq:GLSlim} 
\ee

\begin{figure}[htbp] 
\includegraphics[width=3.2in]{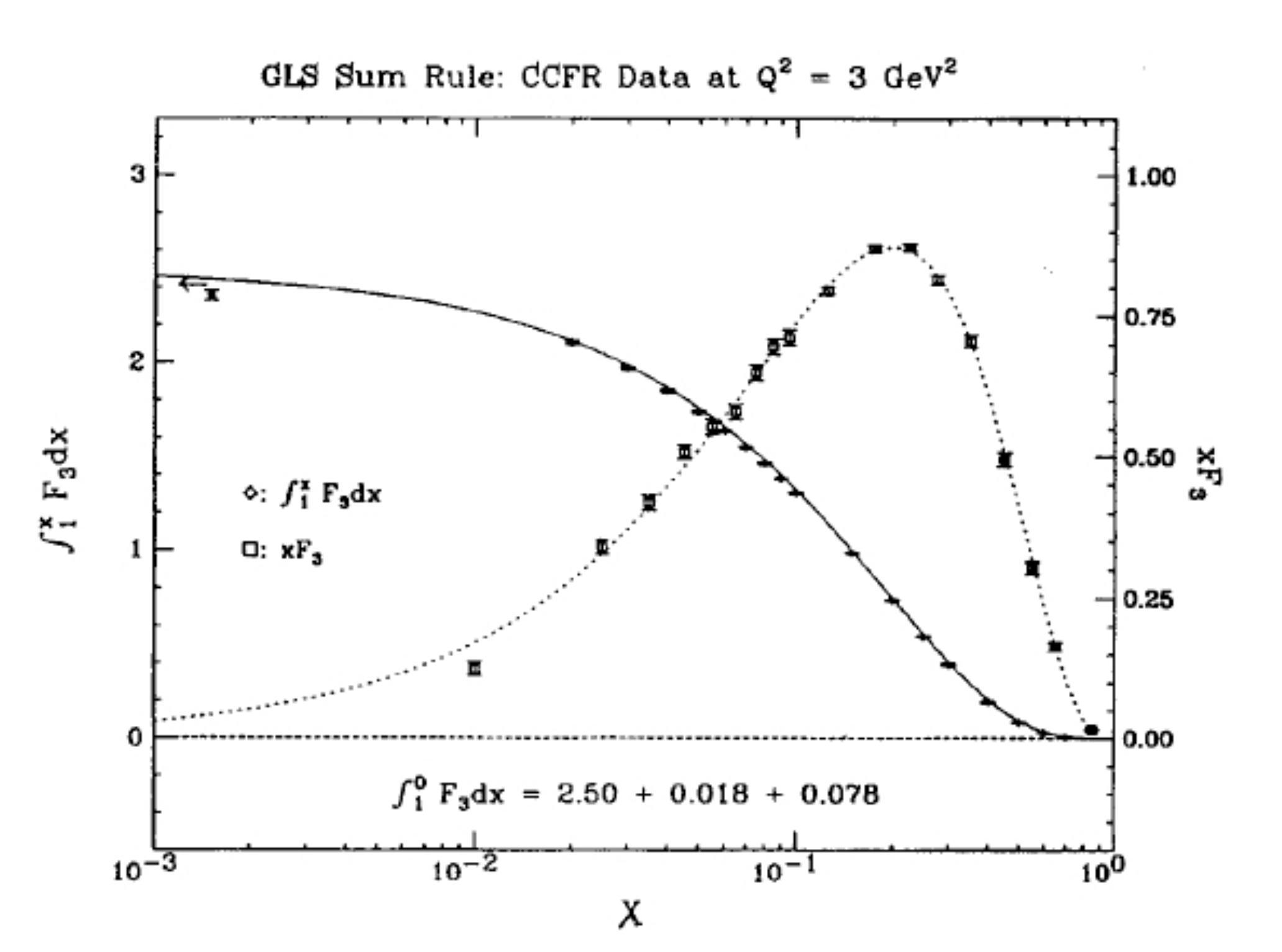} \hspace{0.5cm}
 \caption{Experimental results for Gross-Llewellyn Smith sum rule,
 Eq.\ \protect\ref{SGLSdef}, from the CCFR group, Ref.\ \protect\cite{CCFR}. 
 Squares: $x\overline{F}_3(x)$, sum of neutrinos plus antineutrinos, at 
 $Q^2 = 3$ GeV$^2$. Dashed curve: analytic fit to $x\overline{F}_3$. Diamonds: 
 approximation to the integral $S_{\GLS}(x)$ of Eq.~\protect\ref{eq:SGLSx}. 
 Solid line: fit to the integral $S_{\GLS}(x)$.}
 \label{fig53}
\end{figure}

Fig.~\ref{fig53} shows the CCFR measurements on iron and the 
experimental values of $x\overline{F}_3^{W A}(x)$ (the sum of the nuclear 
$xF_3$ structure function for neutrinos plus that for antineutrinos) vs.~$x$.  
The CCFR group measured cross sections at several values of 
 $x$ and $Q^2$. The squares give the value of $x\overline{F}_3(x)$ 
 interpolated to an average momentum 
 transfer $Q^2 = 3$ GeV$^2$ (this is the mean $Q^2$ for the lowest $x$-bin 
 in the CCFR experiment, since the lowest $x$ values contribute the 
 greatest amount to the GLS sum rule). The dashed curve is the best fit to 
 $x\overline{F}_3(x)$ of the form $Ax^b(1-x)^c$. This form was used to 
 extrapolate the first moment to $x = 0$. The CCFR reported value for 
 the sum rule \cite{CCFR} at 
 this $Q^2$ value is $S_{GLS} = 2.50 \pm 0.018 \ (stat) \pm 0.078 \ (syst)$.  
 The GLS sum rule is therefore known to about 3\%. 

The solid curve in Fig.~\ref{fig53} is $S_{\GLS}(x)$. In the following 
sections we will consider additional QCD contributions. 
We will estimate each correction term as a function of $x$. The lowest $x$ 
value contributing to the Gross-Llewellyn Smith sum rule as measured by 
the CCFR group is 
$x_{min} = 0.015$. We will 
calculate each contribution to the GLS sum rule as a function of $x$, and 
estimate the contribution $\delta S_{\GLS}(x)|_{x=x_{min}}$.   

 \begin{figure}[ht]
\includegraphics[width=2.7in]{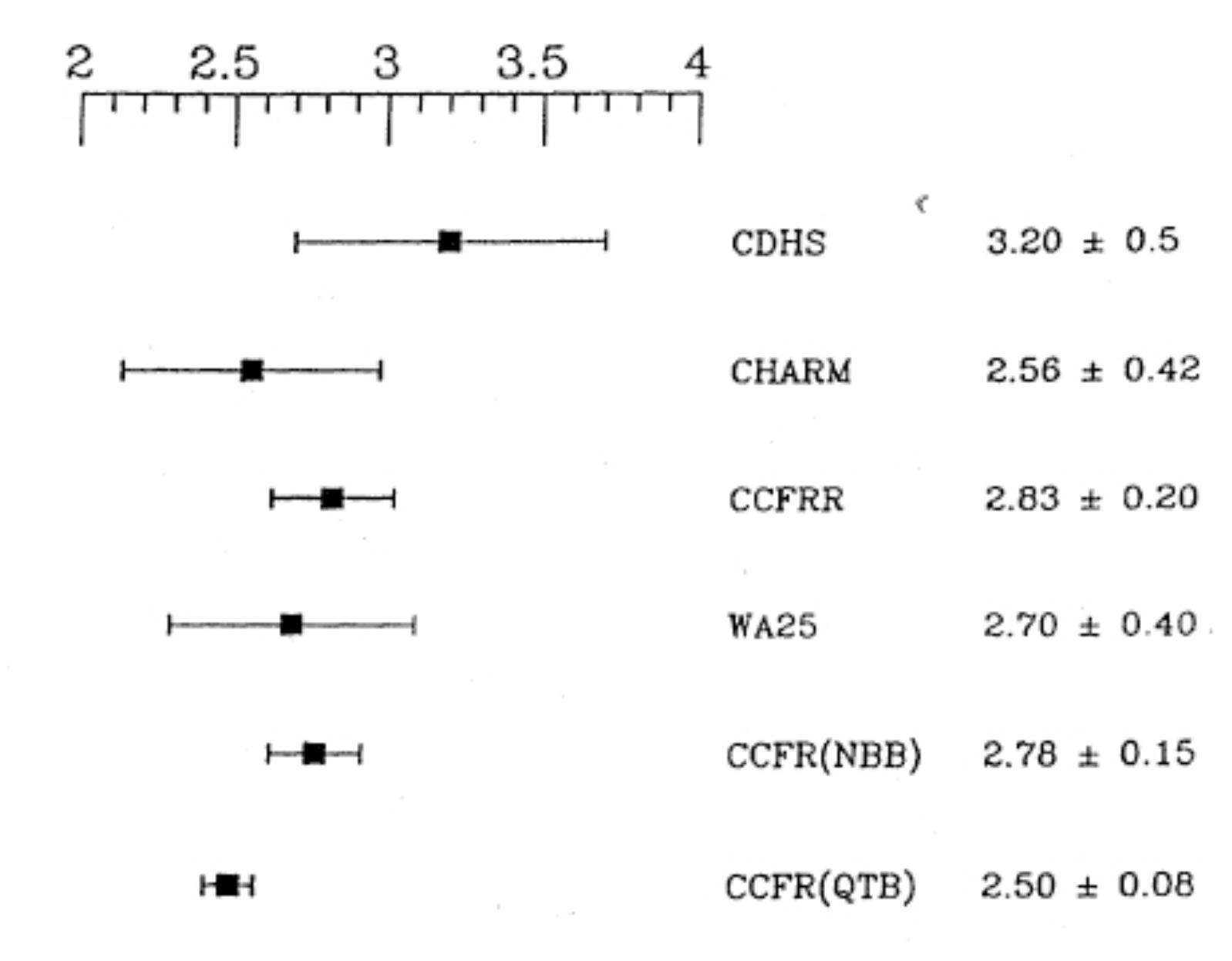}
 \caption{Gross-Llewellyn Smith sum rule, and errors, for a series of 
experiments, in chronological order from top to bottom.}
 \vspace{0.1truein}
 \label{Fig:chrono}
 \end{figure}

Fig.~\ref{Fig:chrono} shows the 
 evolution over time of the GLS sum rule value. The measurements shown 
 are from the CDHS \cite{CDHS2}, CHARM \cite{CHARM}, CCFRR \cite{Macfar}, 
 and WA25 \cite{WA25} collaborations. There are also two points from 
 the CCFR measurements, the first using the Narrow Band Beam (NBB) 
 neutrino data \cite{NBB,Mishra} and the second using the QTB 
 data \cite{CCFR} from the Fermilab Tevatron.  

\begin{figure}[ht]
\includegraphics[width=3.0in]{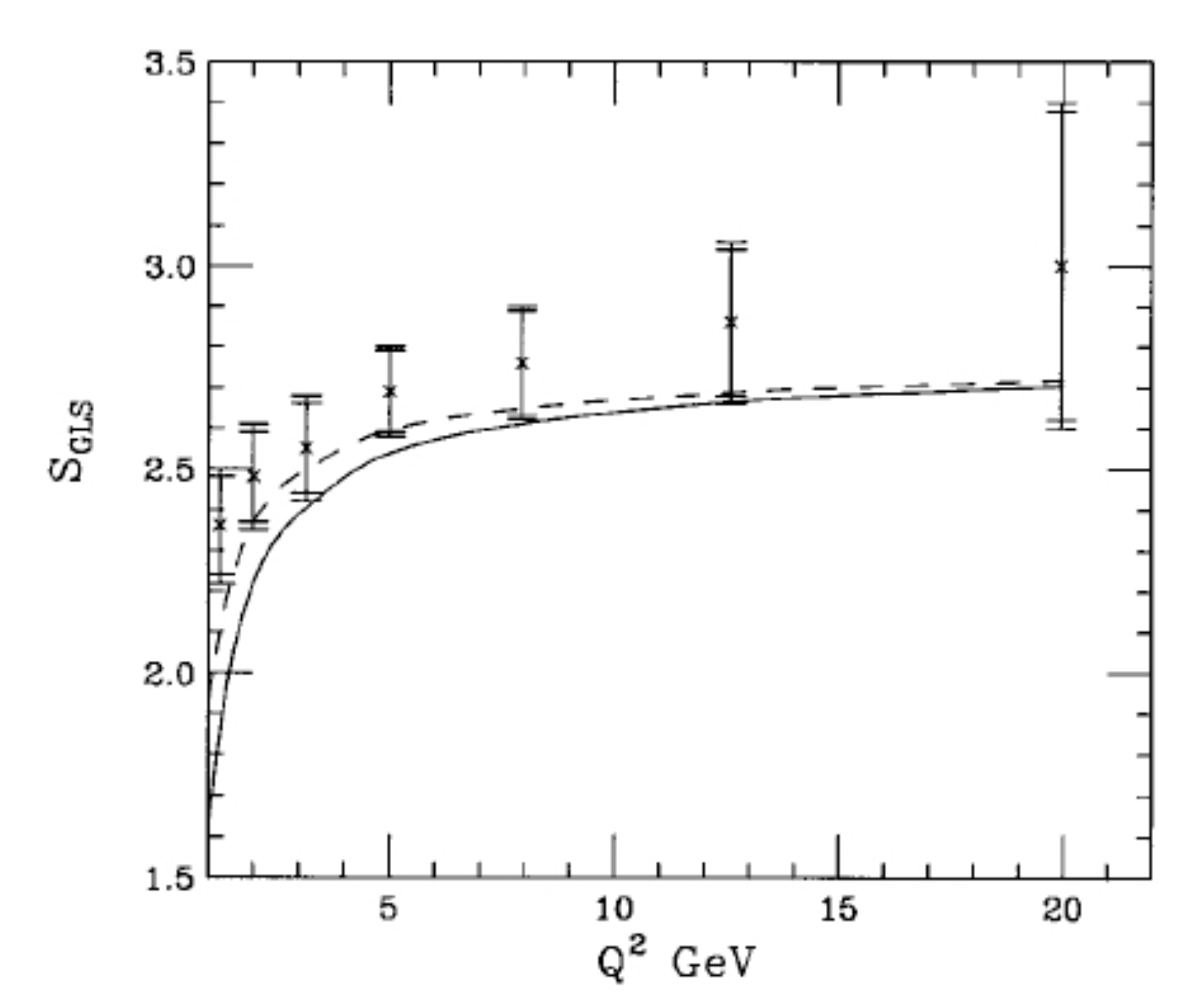}
 \caption{Results for the Gross-Llewellyn Smith sum rule, $S_{\GLS}$ for 
various values of $Q^2$; from Ref.~\protect\cite{Hin96}. The inner error bars are 
statistical and the outer errors combine statistical and systematic errors. The 
data are from the CCFR experiment \protect\cite{CCFR}, and the solid and dashed 
curves are theoreteical QCD predictions from Hinchliffe and Kwiatkowski 
\protect\cite{Hin96}; the solid curve includes higher-twist effects while 
the dashed curve neglects them.}
 \vspace{0.1truein}
 \label{Fig:GLSQ2}
 \end{figure}

The points with error bars in Fig.~\ref{Fig:GLSQ2} represent the experimental 
results from the CCFR group for the  
Gross-Llewellyn Smith sum rule as a function of $Q^2$. The curves are 
theoretical QCD predictions by Hinchliffe and Kwiatkowski \cite{Hin96}, using 
higher-order QCD corrections from Larin and Vermaseren \cite{larin}, and 
higher-twist corrections of Braun and Kolesnichenko \cite{braun}. The dashed 
curves are calculations without higher-twist effects, and the solid curves 
include higher-twist. The theoretical calculations appear to lie 
systematically below the experimental results by one to two standard 
deviations.  

In the remainder of this paper we will review the steps that are taken to 
extract the $F_3$ structure functions from the experimental cross sections. 
We will then review the corrections to Eq.~(\ref{eq:GLSnaive}). In
particular, we will focus on the contributions to the Gross-Llewellyn Smith 
sum rule from strange quarks and from charge symmetry violating contributions 
to parton distribution functions. Although these constitute fairly small 
corrections, nevertheless they may play an important role in determining 
the extent of the agreement, or disagreement, between theory and experiment 
for the GLS sum rule.

 \section{Extracting Structure Functions from Neutrino and Antineutrino 
 Cross Sections}
 \label{Sec:Extract}

The most precise neutrino and antineutrino DIS cross sections are those of 
the CCFR and NuTeV groups, both taken on iron targets. From 
Eq.~(\ref{eq:sig2CC}) by  
taking sums and differences of differential cross sections 
for charged-current DIS from neutrino and antineutrino beams, we can isolate 
different combinations of structure functions. The CCFR and NuTeV 
experiments bin the data in $x$ and $Q^2$. Defining the quantity 
$c = \pi/ \gFsq M E_{\nu}$, it is straightforward to show that 
\bea
&\,&c \left[ \frac{d^2\sigma^{\nu A}_{\CC}}{dx\,dQ^2} + 
  \frac{d^2\sigma^{\ovnu A}_{\CC}}{dx\,dQ^2} \right] = 
  2f(y,Q^2)\overline{F}_2^{W A}(x) \nonumber \\ &+& 
  g(y,Q^2) \Delta xF_3^{W A}(x) \ ; \nonumber \\ 
 &\,&c \left[ \frac{d^2\sigma^{\nu A}_{\CC}}{dx\,dQ^2} - 
  \frac{d^2\sigma^{\ovnu A}_{\CC}}{dx\,Q^2} \right] = 
 f(y,Q^2)\Delta F_2^{W A}(x) \nonumber \\ &+& 
  2g(y,Q^2) x\overline{F}_3^{W A}(x) \ .
\label{eq:Xsumdiff}
\eea
In Eq.~(\ref{eq:Xsumdiff}) the coefficients $f(y,Q^2)$ and $g(y,Q^2)$ are 
defined as 
\bea 
  &\,&f(y,Q^2) = \frac{yf_1(y)}{Q^2} = (1-y+ \frac{y^2}{2(1+ R_L^{\nu})})
  \frac{y}{Q^2}; \nonumber \\ 
&\,&   g(y,Q^2) = \frac{yf_2(y)}{Q^2} = (y - \frac{y^2}{2})\frac{y}{Q^2}. 
\label{eq:fgDef}
\eea 
The cross sections entering Eq.~(\ref{eq:Xsumdiff}) have been 
normalized to give the correct total 
cross sections for neutrinos and antineutrinos. The procedure for this is 
described in the review by Conrad \EA~\cite{Con98}. For the time being, 
we will consider the extraction of the structure function $xF_3^{W A}$ for 
an isoscalar target.  

Previous analyses have neglected the term $\Delta F_2^{W A}(x)$ in 
Eq.~(\ref{eq:Xsumdiff}). From Eq.~(\ref{eq:FApdf}) we see that all 
contributions to this term should be 
small. We will neglect the $c^-$ term in Eq.~(\ref{eq:GLSnaive}) because,
even though a mechanism has been identified which could produce such 
an asymmetry~\cite{Melnitchouk:1997ig}, the charm contribution is 
certainly suppressed substantially with respect to that
associated with strange quarks and the charmed contribution is also 
kinematically suppressed  for the experimental conditions -- c.f. 
Eq.~(\ref{eq:sigcorr}) in Sect.~\ref{Sec:GLSnuc}. Thus for an isoscalar target 
at sufficiently high $Q^2$ we expect 
\bea  
\Delta F_2^{W A}(x) &\to& 2xs^-(x) + x(-\delta u^-(x) + 
  \delta d^-(x)); \nonumber \\ 
\Delta xF_3^{W A}(x) &\to& 2x(s^+(x)+ c^+(x)) \nonumber \\ 
 &+& x(-\delta u^+(x) + \delta d^+(x)). \nonumber \\ 
\label{eq:FApdfN}
\eea
From Eq.~(\ref{eq:FApdfN}) we see that for an isoscalar target the term 
$\Delta F_2^{W A}(x)$ will be non-zero only if one has a strange quark 
momentum asymmetry $s^-(x) \ne 0$, and/or non-zero valence quark CSV 
contributions. In Sect.~\ref{Sec:GLSnuc} we will discuss additional 
contributions to $\Delta F_2^{W A}(x)$ for a non-isoscalar target. 

If one neglects the term $\Delta F_2^{W A}(x)$ in Eq.~(\ref{eq:Xsumdiff}) 
(as was the case in the analysis of the CCFR data), then the structure 
function $x\overline{F}_3^{W A}(x)$ will just be 
proportional to the difference between the neutrino and antineutrino 
charged-current DIS cross sections. For a given $x$ bin, the structure 
function $xF_3$ is then given by averaging the structure function 
differences over the $Q^2$ bin appropriate to the given $x$ bin. Thus 
we obtain 
\bea
x\overline{F}_3^{W A}(x) &=& \frac{c}{2A(x)}\int_{\langle Q^2\rangle} \left[ 
  \frac{d^2\sigma^{\nu A}_{\CC}}{dx\,dQ^2} - 
  \frac{d^2\sigma^{\ovnu A}_{\CC}}{dx\,Q^2} \right]\, dQ^2, 
  \nonumber \\ A(x) &=& \int_{\langle Q^2\rangle} g(y,Q^2) dQ^2 \ .  
\label{eq:intQ2}
\eea
In Eq.~(\ref{eq:intQ2}), $\langle Q^2 \rangle$ denotes the average over 
the $Q^2$ bin appropriate to a given $x$ bin. In Eq.~(\ref{eq:intQ2}), 
we have neglected the slow variation of $\overline{F}_3^{W A}$ with $Q^2$. 

However, if the quantity $\Delta F_2^{W A}(x)$ is non-zero, then 
Eq.~(\ref{eq:intQ2}) will not give the structure function 
$x\overline{F}_3^{W A}(x)$, but rather a linear combination of 
$xF_3$ and $\Delta F_2$. Comparing this with Eq.~(\ref{eq:Xsumdiff}), we 
note that the $y$ dependence of the coefficients $f$ and $g$ of the two 
terms is quite different, as shown in Eq.~(\ref{eq:fgDef}). In particular, 
the coefficient of $x\overline{F}_3^{W A}$ vanishes 
at $y=0$, while the coefficient of $\Delta F_2^{W A}$ is finite. Inserting 
Eq.~(\ref{eq:Xsumdiff}) into 
Eq.~(\ref{eq:intQ2}) and averaging over the $Q^2$ range for each $x$ bin gives 
\bea
 &\,& \frac{c}{2A(x)}\int_{\langle Q^2\rangle} \left[ 
  \frac{d^2\sigma^{\nu A}_{\CC}}{dx\,dQ^2} - 
  \frac{d^2\sigma^{\ovnu A}_{\CC}}{dx\,Q^2} \right]\, dQ^2 \nonumber \\  
 &=& x\overline{F}_3^{W A}(x) + B(x)\Delta F_2^{W A}(x),  
  \nonumber \\ &\,&B(x) = \frac{\int_{\langle Q^2\rangle} f(y,Q^2) dQ^2}
  {2A(x)} \ .  \label{eq:intcorrect}
\eea

\begin{figure}[ht]
\includegraphics[width=2.9in]{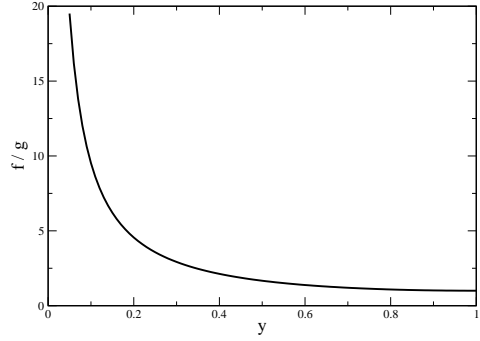}
 \caption{The quantity $f(y,Q^2)/g(y,Q^2)$ from Eq.~(\protect\ref{eq:Xsumdiff})
plotted versus $y$, under the assumption that $R_L^{\nu} = 0$; it is 
identical to $f_1(y)/f_2(y)$ defined in Eq.~(\protect\ref{eq:sig2CC}).} 
 \vspace{0.1truein}
 \label{Fig:frat}
 \end{figure}

The quantity $B(x)$ in Eq.~(\ref{eq:intcorrect}) is the relative weighting 
between the $\Delta F_2$ and $xF_3$ terms. $B(x)$ will depend upon the $x$ bin 
and the $Q^2$ values that are averaged over for each $x$ bin. In 
Fig.~\ref{Fig:frat} we plot the ratio $f(y,Q^2)/g(y,Q^2)$ 
in Eq.~(\ref{eq:Xsumdiff}) vs.~$y$ (this quantity is identical 
to the ratio $f_1(y)/f_2(y)$ from Eq.~(\ref{eq:sig2CC})). This ratio is always 
greater than one, and becomes quite large at small $y$ values. The quantity 
$B(x)$ is normalized to equal one if one integrates over all $y$; however 
$B(x)$ could be greater than one particularly if the average  
over $Q^2$ is weighted towards small $y$ values. 

Eq.~(\ref{eq:intcorrect}) shows that for an isoscalar target, the common 
process of taking the 
difference of neutrino and antineutrino charged-current cross sections, 
and averaging over the $Q^2$ bin for a given $x$ bin produces  
a linear combination of 
$x\overline{F}_3^{W A}$ and $\Delta F_2^{W A}$ with a relative weighting   
$B(x)$. From Eq.~(\ref{eq:FApdf}) we see that the partonic content of   
the quantity identified as $x\overline{F}_3$ will be 
\bea
&\,&''x\overline{F}_3^{W A}(x)'' = x\overline{F}_3^{W A}(x) + 
  B(x)\Delta F_2^{W A}(x) \nonumber \\ &=& x\bigl[ u^-(x) + d^-(x) + 
  s^-(x)(1 + 2B(x)) \nonumber \\ &+& \delta d^-(x)(B(x) - \frac{1}{2}) - 
  \delta u^-(x)(B(x) + \frac{1}{2}) \bigr].  \nonumber \\ 
\label{eq:F3extract}
\eea
We put quotes around the quantity $x\overline{F}_3^{W A}$, since it 
represents the linear combination of $x\overline{F}_3^{W A}$ and 
$\Delta F_2^{W A}$ obtained for that $x$ bin. As we have mentioned, 
the $y$ dependence of the coefficients 
$f(y,Q^2)$ and $g(y,Q^2)$ in Eq.~(\ref{eq:Xsumdiff}) is quite different, 
particularly in the forward direction. When the longitudinal to transverse 
ratio $R_L^{\nu} = 0$, the ratio 
$f(y,Q^2)/g(y,Q^2)$ is given 
in Fig.~\ref{Fig:frat}. If we assume that for each $x$ bin, 
the data is averaged over all $y$, then one obtains $B(x) = 1$, 
and Eq.~(\ref{eq:F3extract}) becomes 
\bea 
&\,& ''x\overline{F}_3^{W A}(x)''|_{B(x)=1} = x\biggl[ u^-(x) + d^-(x)  
 \nonumber \\ &+& 3s^-(x) + \frac{\delta d^-(x) - 3\delta u^-(x)}{2} \biggr]. 
  \nonumber \\ \label{eq:averagey}
\eea 

These additional terms should be included in the analysis of the 
experimental data. In the absence of such an analysis we will provide 
estimates of the sign and magnitude of these corrections and their 
effect on the Gross-Llewellyn Smith sum rule. 

\section{Additional Corrections to the GLS Sum Rule}
\label{Sec:sCSV}

 In addition to contributions from the light valence quarks, higher order 
QCD terms and higher twist contributions, Eq.~(\ref{SGLSdef}) contains 
additional QCD corrections. These corrections are included in the experimental 
determination of the GLS sum rule, but these additional terms have not been 
included in theoretical calculations. There are two types of corrections. The 
first involves additional contributions to 
the desired term $x\overline{F}_3^{W A}$ in Eq.~(\ref{eq:FApdf}). The second 
contribution appears by virtue of the term $\Delta F_2^{W A}$, which has 
not been separated from the desired term. In both cases we include 
contributions from strange quarks and partonic CSV corrections. Although the 
first moment of each of these contributions vanishes when integrated over all  
$x$, there is a residual contribution still present at 
$x_{min} = 0.015$, the lowest data point in the CCFR experiment. In order to 
reconcile theoretical and experimental determination of the GLS sum rule, we 
choose to subtract these additional contributions from the experimental data 
when comparing with theory. We define the 
corrections to the GLS sum rule as
\be 
\delta S_{\GLS} = {\rm lim}_{x \rightarrow 0}\ \delta S_{\GLS}(x). 
\label{eq:corrdef}
\ee
For an isoscalar target these corrections have the form  
\bea
\delta S_{\GLS}(x) &=& f(\als)\left[ \delta S_{\GLS}^s(x) + 
\delta S_{\GLS}^{\CSV}(x)\right]  \ , 
  \nonumber \\ \delta S_{\GLS}^s(x) &=& \int_x^1 s^-(y)(2B(y)+1) \, dy \ , 
  \nonumber \\ 
\delta S_{\GLS}^{\CSV}(x) &=& \int_x^1 \biggl[ (B(y)-\frac{1}{2})\delta d^-(y) 
  \nonumber \\ &-& (B(y)+\frac{1}{2})\delta u^-(y)\biggr] \, dy 
\label{eq:GLSscsv}
\eea
The term $f(\als)$ in Eq.~(\ref{eq:GLSscsv}) is the QCD correction that 
has been calculated by Larin and Vermaseren \cite{larin}; it is the 
term in square brackets in Eq.~(\ref{SGLSdef}).  

Eq.~(\ref{eq:GLSscsv}) contains two terms. The first is the contribution 
from the strange quark asymmetry. The second is the contribution from charge 
symmetry violating valence 
PDFs. An additional effect will result from the nuclear modification of the 
parton distributions. Implicitly, all of the parton distribution functions 
in Eq.~(\ref{eq:GLSscsv}) denote parton distributions in iron. In Sect. 
\ref{Sec:GLSnuc} we will discuss nuclear modifications of the PDFs. Note that 
the terms containing the quantity $B(y)$ result from the contamination of 
the $x\overline{F}_3$ structure function from the $\Delta F_2$ term. 

If the quantity $B(y)$ in Eq.~(\ref{eq:GLSscsv}) was a constant, then 
both the strange and CSV terms would give zero in the limit $x \rightarrow 0$. 
This is because valence quark normalization requires that the valence strange 
quark and valence CSV PDFs have zero first moment. However, we have no 
reason to believe that $B(y)$ will be constant. Note also that the $C$-odd 
strange quark and valence CSV effects contribute to $\delta S_{\GLS}(x)$ at 
any finite value of $x$. So even if the quantity $B(y)$ were a constant, the 
CSV and strange quark effects would be finite for any non-zero value of $x$, 
vanishing only at $x=0$.  

From our current understanding of the parton distributions, for sufficiently 
large $x$ we expect every term in the quantity $\Delta F_2^{W A}(x)$ in 
Eq.~(\ref{eq:FApdfN}) to have the same sign. As we shall see, all of the 
latest analyses of strange quark distributions \cite{Mason:2007zz,Lai:2007dq,Martin:2009iq,Ball:2009mk,Alekhin:2008mb} find that the quantity $s^-(x)$ is positive for sufficiently 
large $x$. Analyses of parton valence charge symmetry violating effects for 
parton distributions \cite{Londergan:2009kj} obtain a quantity 
$\delta d^-(x)- \delta u^-(x) > 0$ for $x \geq 0.1$. Consequently, both of the 
terms in $\delta S_{\GLS}$ in Eq.~(\ref{eq:GLSscsv}) will contribute with the 
same sign. In the following sections we will estimate the magnitude of each of 
these contributions. 

\subsection{Corrections to the GLS Sum Rule at $Q^2 = 8$ GeV$^2$}
\label{Sec:GLScorrs}

For convenience we calculate the corrections associated with a strange 
quark asymmetry and parton charge symmetry violation at a single value of 
$Q^2$, and we   
choose $Q^2 = 8$ GeV$^2$. From Fig.~\ref{Fig:GLSQ2} we see that this is 
a value of $Q^2$ for which the experimental GLS sum rule has been 
measured. It is also a relatively convenient value of $Q^2$ for which 
to estimate corrections from both strange quarks and parton CSV. At 
$Q^2 = 8$ GeV$^2$, the experimental value of the Gross-Llewellyn Smith 
sum rule is $S_{\GLS}^{exp} = 2.76 \pm 0.14$, and the theoretical value 
of Hinchliffe and Kwiatkowski \cite{Hin96} including higher-twist corrections 
is $S_{\GLS}^{th} = 2.62$. So the theoretical value is just over one standard 
deviation below the experimental result. 

In the next sections we will provide estimates for the strange quark and 
partonic CSV contributions to the GLS sum rule. After evaluating them we 
will determine the correction that they make to the experimental value and 
error for the Gross-Llewellyn Smith sum rule for this value of $Q^2$. 

The QCD correction that appears in Eq.~(\ref{eq:GLSscsv}) was evaluated by 
Larin and Vermaseren \cite{larin}; it has the form  
\be 
f(\als) = 1- \frac{\als}{\pi} - \left(\frac{\als}{\pi}\right)^2 r_1 -  
\left(\frac{\als}{\pi}\right)^3 r_2 \ , 
\label{eq:fcorr}
\ee
where for $n_f = 3$ active flavors one has $r_1 = 3.5833$ and $r_2 = 18.9757$ 
\cite{larin}. For the strong coupling $\als$ we use the value chosen by 
Hinchliffe and Kwiatkowski \cite{Hin96} in their theoretical calculations. 
In the $\MSb$ factorization scheme they chose the scale parameter 
$\Lambda^{(5)}_{\MSb} = 233$ MeV; here the superscript denotes $n_f$. 
This corresponds to $\Lambda^{(3)}_{\MSb} = 366$ MeV and a strong coupling 
$\als(Q^2 = 3~GeV^2) = 0.336$. Using this value we then obtain 
$f(\als)(Q^2 = 8~GeV^2) = 0.885$. 

\subsection{Contributions from Strange Quarks}
\label{Sec:GLSstrange}

The strange quark parton distributions are best obtained from an analysis of 
opposite-sign dimuon production in reactions induced by neutrinos and 
antineutrinos. In such reactions, dimuon production from a $\nu$ ($\ovnu$) 
beam is sensitive to the $s$ ($\bar{s}$) distribution, so that in principle 
a comparison of these cross sections could enable one to determine differences 
between the $s$ and $\bar{s}$ PDFs.  There are recent measurements of these 
reactions by the CCFR and NuTeV \cite{Baz95,Gon01,Mason:2007zz} collaborations. In 
the CCFR experiment the $\nu$ and $\ovnu$ beams were not separated and the 
type of reaction was inferred from the charge of the faster muon, while 
 the NuTeV experiment used separated $\nu$ and $\ovnu$ beams.The 
correction to the GLS sum rule is obtained from 
\be 
\delta S_{\GLS}^s(x) = \int_x^1 s^-(y)(2B(y)+1) \, dy 
\label{eq:Scontrib}
\ee

Now, the first moment of $s^-$ is zero, from valence quark normalization 
(there are no net ``strange valence'' quarks in the nucleon). However, 
recent phenomenological analyses of strange quark distributions all obtain 
qualitatively similar results. All of them find the most probable value 
is a positive strange quark momentum asymmetry, $\langle xs^-(x)\rangle > 
0$. Also, the best fit to the quantity $s^-(x)$ changes sign at an extremely 
small value of $x$ and is large and positive down to rather small $x$ values.  

 \begin{figure}[ht]
\includegraphics[width=3.0in]{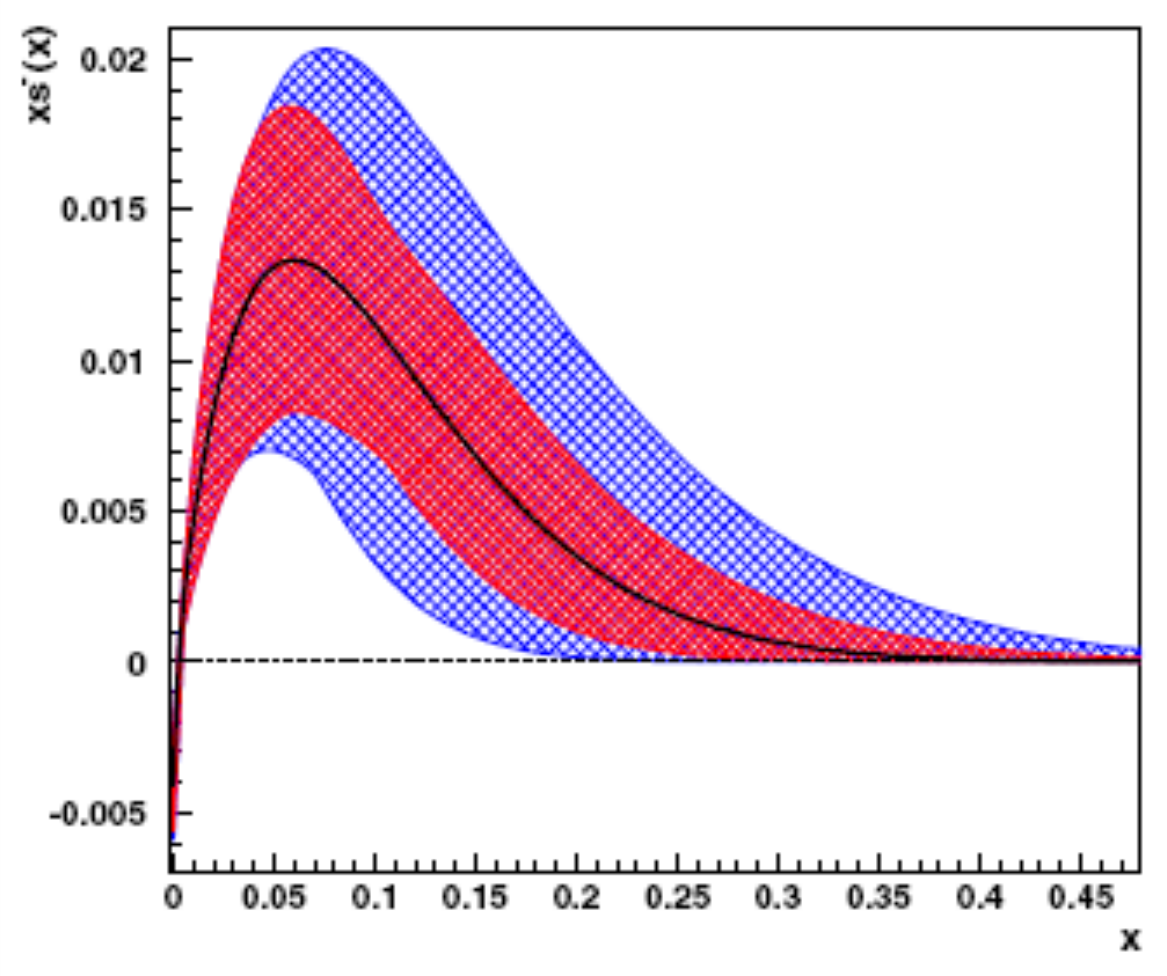}
 \vspace{0.15truein}
 \caption{[color online] The quantity $xs^-(x) = x[s(x) - \bar{s}(x)]$, 
 vs.~$x$, as extracted by the NuTeV Collaboration, 
 Ref.~\protect\cite{Mason:2007zz}. 
 Values are obtained for $Q^2 = 16$ GeV$^2$. The outer error band is the 
 combined error, while the inner band is without the uncertainty in the 
 semileptonic branching ratio $B_c$.} 
 \vspace{0.1truein}
 \label{Fig:Mason}
 \end{figure}

For example, the analysis by Mason \EA~\cite{Mason:2007zz} obtains a best value 
for the the integral of $xs^-(x)$ 
\bea 
  S^- &=& 0.00196 \pm 0.00046(stat) \pm 0.00045(syst) \nonumber \\ 
   &\,& ^{+0.00148}_{-0.00107}(external) .  
\label{eq:Sv}
\eea
The quantity $Q$ refers to the second moment of a parton distribution $q(x)$, 
\IE 
\be 
Q \equiv \int_0^1 x\, q(x)\, dx 
\label{eq:Qdef}
\ee
 The Mason result is obtained for a value $Q^2 = 16$ GeV$^2$. In 
Eq.~(\ref{eq:Sv}), the term ``external'' refers to the contribution 
arising from uncertainties on external measurements.  

Fig.~\ref{Fig:Mason} plots the quantity $xs^-(x)$ vs.~$x$ from the latest 
NuTeV analysis. The strange quark momentum asymmetry, $S^-$, is quite 
sensitive to two quantities. The first is the semileptonic branching ratio 
$B_c$; the outer band in Fig.~\ref{Fig:Mason} shows the result for $S^-$ with 
the $B_c$ uncertainty, and the inner band is the result without the $B_c$ 
uncertainty. The second is the point at which the distribution $xs^-(x)$ 
crosses zero (it must cross zero at least once to give zero first moment for 
$s^-(x)$). The current best fit crosses zero at a very small value $x \sim 
0.004$. This means that the quantity $s^-(x)$ would have a large negative 
spike at extremely low $x$ (in fact, an $x$ value smaller than the lowest $x$ 
point measured in the experiment). 

\begin{table*}[btp]
\begin{ruledtabular}
\begin{tabular}{lccc}
    & $S^- = \langle xs^-\rangle$  & $x_0$ & $Q^2$ (GeV$^2$) \\
\hline
Mason et al.~\cite{Mason:2007zz}     
& $0.00196 \pm 0.00143$  & $0.004$  & $16$ \\
CTEQ~\cite{Lai:2007dq} 
& $0.0018_{-0.0004}^{+0.0016}$ & $0.01-0.02$ & $1.69$ \\
NNPDF~\cite{Ball:2009mk}             
& $0.0005 \pm 0.0086$    & $0.13$  & $20$  \\
MSTW~\cite{Martin:2009iq}            
& $0.0016_{-0.0009}^{+0.0011}$ & $0.016$ & $1.0$ \\
Alekhin et al.~\cite{Alekhin:2008mb} 
& $0.0013 \pm 0.0009 \pm 0.0002$ & $\leq 0.02$ & $8$ 
\end{tabular}
\end{ruledtabular}
\caption{A summary of recent phenomenological estimates of the strangeness 
asymmetry ($S^-$ ), the crossover point $x_0$ and the $Q^2$ value at which the 
asymmetry is obtained. }
\label{Tab1}
\end{table*}

There are several other recent phenomenological estimates of the strange quark 
asymmetry. All are very sensitive to the CCFR and NuTeV dimuon 
production data. The CTEQ group \cite{Lai:2007dq} obtains 
$S^- = 0.0018^{+0.0016}_{-0.0004}$ at their starting scale $Q_0^2 = 1.69$ GeV$^2$. 
Their best fits to $xs^-(x)$ found crossovers in the 
vicinity $x_0 \sim 0.01-0.02$. The NNPDF Collaboration \cite{Ball:2009mk} used 
only the NuTeV data and not the CCFR results. They report a value $S^- = 0.0005 
\pm 0.0086$ at $Q^2 = 20$ GeV$^2$; the exceptionally large error in the NNPDF 
value (a factor of 5 to 6 larger than the errors from the other analyses) 
results in part from their use of a neural network procedure, which 
does not build in widely accepted constraints on the shape of sea quark 
distributions.  

Two other groups supplemented the NuTeV and CCFR dimuon data with charm 
production data from CHORUS \cite{KayisTopaksu:2005je,Onengut:2005kv}, 
which helps to constrain the semileptonic branching ratio. The MSTW group 
\cite{Martin:2009iq} obtains $S^- = 0.0016^{+0.0011}_{-0.0009}$ at $Q^2 = 10$ 
GeV$^2$; their next to leading order (NLO) fit had a crossover 
$x_0 = 0.016$ at the starting scale $Q_0^2 = 1$ GeV$^2$. Alekhin, Kulagin 
and Petti \cite{Alekhin:2008mb} obtained $S^- = 0.0013 \pm 0.0009 
\pm 0.0002$ at $Q^2 = 8$ GeV$^2$, and their best fit to $xs^-(x)$ had a 
crossover $x_0 \leq 0.02$. For these various phenomenological fits we 
summarize the values of $S^-$, the crossover point $x_0$ and the value 
of $Q^2$ at which the asymmetry is calculated in Table I. 

From Eq.~(\ref{eq:Scontrib}) the contribution to the GLS sum rule from strange 
quarks, for a given value of $x$, will be given by the integral of $s^-(y)$ 
weighted by the quantity $2B(y)+1$. If we approximate $B(y)$ as a constant 
we just need the integral of $s^-(y)$ from $x$ to $1$. We choose 
the smallest value $x_{min}= 0.015$ measured by CCFR. To estimate the integral  
we made an analytic fit to the strange asymmetry measured by Mason \EA, 
in the region $ x \ge 0.004$ \cite{Mason:2007zz}. Our fit had the form 
$s^-(x) = a x^b e^{-cx} (x- 0.004)$. With this fit we obtain the result   
\be 
\int_{x=0.015}^1 s^-(y)dy \approx 0.026.
\label{eq:xsmanal}
\ee
Within their error bars, all of the phenomenological fits now obtain a positive 
value for the strange quark momentum asymmetry. Because of their unusually large 
error on the strange quark asymmetry and its unphysically large value in the 
valence region, we do not use the NNPDF 
result. All of the other fits to the strange quark PDFs produce a quantity 
$xs^-(x)$ which changes sign at an extremely small value of $x$. Thus all 
of these strange quark PDFs will give a reasonably large contribution to the 
GLS sum rule at a value of $x$ corresponding to the lowest $x$ value 
measured in the CCFR experiment. We assign an error of 75\% which is 
roughly the average of the four determinations (excluding NNPDF) summarized 
in Table I. Thus we choose 
\be 
\int_{x=0.015}^1 s^-(x)dx = 0.031 \pm 0.023.
\label{eq:xsmcorr}
\ee

In Eq.~(\ref{eq:xsmcorr}) we have increased the integral of this 
distribution by 20\%; this represents the approximate increase in this 
moment in evolving from $Q^2 = 16$ GeV$^2$ to the value $Q^2 = 8$ GeV$^2$ 
appropriate for our evaluation of the GLS sum rule. This increase is 
comparable to results obtained by the NNPDF group \cite{Ball:2009mk}, who 
performed DGLAP evolution on the second moment of strange quark distributions 
to extrapolate in $Q^2$.   

From Eq.~(\ref{eq:Scontrib}), the strange quark asymmetry contribution 
to the GLS sum rule will depend on estimates of $B(x)$. We make two 
simple guesses for this quantity. First, we take $B(x)=1$; this is the 
result if the integral of Eq.~(\ref{eq:intcorrect}) was taken over all 
$y$. Next, we estimate $B(x) = 2$; this represents an upper limit (and 
possibly an overestimate) of this quantity. Under these approximations we 
obtain strange corrections to the GLS sum rule  
\bea 
\delta S_{\GLS}^s &=& 0.093 \pm 0.063 \ ,~{\rm for}~B(x) = 1 \ ; \nonumber \\   
\delta S_{\GLS}^s &=& 0.156 \pm 0.115 \ ,~{\rm for}~B(x) = 2 \ .
\label{eq:Sgls}
\eea
The large result for the strange quark contribution results from the 
fact that the strange quark momentum asymmetry changes sign at an extremely 
small value of $x$. Although we have used the result of Mason \EA, this 
property is shared by all phenomenological strange quark analyses in Table I 
except for NNPDF. 

The contribution from the strange momentum asymmetry to the GLS sum rule is 
strongly dependent on the crossover point $x_0$ at which $s^-(x)$ crosses zero. 
If the crossover point for $s^-(x)$ occurred at a value $x_0 \geq 0.1$ then 
strange quarks would make an extremely small contribution to the GLS sum rule. It 
is difficult to imagine a physical mechanism that would cause $s^-(x)$ to change 
sign at such small crossover points $x_0$ as have been found in these 
phenomenological analyses 
\cite{Mason:2007zz,Lai:2007dq,Martin:2009iq,Alekhin:2008mb}. Indeed, model 
calculations almost invariably yield a zero at $x \sim 0.1$ or 
higher~\cite{Signal:1987gz,Melnitchouk:1999mv,Carvalho:2005nk,Wei:2007nb}. The 
NuTeV group found that with a moderate increase in $\chi^2$ one could obtain 
considerably larger values of $x_0$ and corresponding large decreases in the 
second moment $S^-$ \cite{Mason:2007zz}.

\subsection{Contributions from Charge Symmetry Violating PDFs}
\label{Sec:GLScsv}

We also can estimate the contribution from parton CSV. For an isoscalar 
target this is given by 
\bea
\delta S_{\GLS}^{\CSV}(x) &=& \int_x^1 \bigl[ (B(y)-\frac{1}{2})\delta d^-(y) 
  \nonumber \\ &-& (B(y)+\frac{1}{2})\delta u^-(y) \bigr] \, dy
\label{eq:CSVcontrib}
\eea
For this we need the valence CSV parton distribution functions 
\cite{Londergan:2009kj}. We adopt the functional form used by the MRST 
group \cite{MRST03}, 
\be
\delta u^-(x) = -\delta d^-(x) = \kappa x^{-0.5}(1-x)^4 (x - .0909)
\label{eq:CSVval}
\ee
The best fit value of MRST was $\kappa = -0.2$. This produced contributions 
very much like the quark model valence CSV calculations of Rodionov 
\EA~\cite{Rod94} evaluated at $Q^2 = 10$ GeV$^2$. Here we choose  
$\kappa = -0.3$, which approximates quite well the quark model 
valence CSV from Rodionov, plus the valence CSV arising from ``QED 
splitting'' \cite{MRST05,Glu05}. If we assume that $\delta d^-(x) = - 
\delta u^-(x)$, then from Eq.~(\ref{SGLSdef}) the term $S_{\GLS}^{\CSV}(x)$
in Eq.~(\ref{eq:CSVcontrib}) has the form 
\be 
\delta S_{\GLS}^{\CSV}(x) = - 2 \, \int_x^1 B(y) \delta u^-(y) \, dy \ .
\label{eq:MRSTform}
\ee 

We insert the analytic form of Eq.~(\ref{eq:CSVval}) into 
Eq.~(\ref{eq:MRSTform}) and evaluate at $x_{min} = 0.015$, the minimum 
$x$ value for the CCFR measurements. As for the strange quark contribution 
we evaluate this using two different values for the weighting function, 
$B(y) = 1$ and $B(y) = 2$. We assign a 100\% error to the CSV contribution. 
Thus we obtain 
\bea 
\delta S_{\GLS}^{\CSV} &=& 0.013 \pm 0.013 \ , {\rm B(y)~=1} 
  \nonumber \\ &=& 0.026 \pm 0.026 \ , {\rm B(y)~=2}
\label{eq:CScontr}
\eea 
 
The partonic charge symmetry violating contributions correspond to a value 
$Q^2 \sim 10$ GeV$^2$. This is sufficiently close to the value $Q^2 = 8$ 
GeV$^2$ that we do not modify this further. 

\subsection{Non-Isoscalar and Nuclear Corrections} 
\label{Sec:GLSnuc}

Until now our equations have assumed an isoscalar target. We must include 
corrections to account for the excess neutrons in iron. From 
Eq.~(\ref{eq:FApdf}) there are a number of small corrections in the 
case that $N \not= Z$. For our purposes the most important will be an 
additional contribution to the quantity $\Delta F_2^{W A}(x)$ of the form 
\be 
\delta (\Delta F_2^{W A}(x)) = \nuf x[u^-(x) - d^-(x)] \ . 
\label{eq:delDF2}
\ee
When multiplied by the quantity $B(x)$, divided by $x$ and integrated 
over all $x$, this leads to the contribution 
\be
\delta S_{\GLS}^{\nuf}(x) = \nuf \,\int_x^1 B(y)[u^-(y) - d^-(y)]\, dy \ . 
\label{eq:delSA}
\ee
For convenience we take the lower limit of this integral to be $x=0$; in 
the limit where $B(x)$ is approximated as a constant, $B(x) \sim B$, the 
remaining integral is just one, so in this approximation we obtain 
\bea
\delta S_{\GLS}^{\nuf} &\approx& \nuf B \ , \nonumber \\ 
    &=& 0.058, \hspace{0.4cm} {\rm for}~B(x) = 1; \nonumber \\ 
    &=& 0.115, \hspace{0.4cm} {\rm for}~B(x) = 2.  
\label{eq:delSAB}
\eea
Note that all of the contributions to the GLS sum rule (strange quark 
asymmetry, quark model valence CSV and QED splitting CSV, and non-isoscalar 
effects) are of the same sign. Thus, their contributions will add 
coherently in modifying the GLS sum rule.  
  
The contributions listed here for strange quarks, CSV parton 
distributions and $N \not= Z$ effects were included in the experimental 
results but not in the theoretical calculations of Hinchliffe and Kwiatkowski 
\cite{Hin96}. Consequently we choose to subtract these contributions from the 
experimental results, in order to compare with theory. At $Q^2 = 8$ GeV$^2$ 
the quoted experimental result for the Gross-Llewellyn Smith sum rule was 
$2.76 \pm 0.14$. We take the strange quark contribution from 
Eq.~(\ref{eq:Sgls}), the CSV contribution from Eq.~(\ref{eq:CScontr}), 
and the $N \not= Z$ contribution from Eq.~(\ref{eq:delSAB}). These effects 
are multiplied by the QCD correction factor $f(\alpha_s)$ from 
Eq.~(\ref{eq:fcorr}). We assume 
that the errors can be combined in quadrature. This leads to the net result  
\bea 
S_{\GLS}^{expt}|_{Q^2 = 8~GeV^2} &\to& 2.62 \pm 0.15, ~{\rm B(y)~=1} 
  \nonumber \\ &\to& 2.50 \pm 0.17, ~{\rm B(y)~=2} \ .\nonumber \\
\label{eq:GLSth}
\eea 
We can compare this with the theoretical value $S_{\GLS}^{th} = 2.62$. We see 
that these additional terms improve the agreement between theory and 
experiment. In the approximation $B(y) =1$ there is now excellent agreement 
between theory and experiment; when $B(y) = 2$ the experimental point is now 
below the data but still within one standard deviation. Since the errors are 
added in quadrature, the net result is a small increase in the overall error.   

For the purpose of completeness we will review the corrections that were 
applied to the experimental data by the CCFR group. Before solving 
for the structure functions from the neutrino and antineutrino cross 
sections (see Eq.~(\ref{eq:Xsumdiff}) and following equations), the 
CCFR collaboration made a series of corrections to the cross sections. First, 
as we have mentioned previously, the neutrino and antineutrino cross sections 
were normalized to the total fluxes. Next, the cross sections were multiplied 
by four nuclear correction factors, 
\be 
\sigma^{corr} = \sigma^{iso}\times \sigma^{rad}\times \sigma^{c}\times 
  \sigma^{W} \ . 
\label{eq:sigcorr}
\ee
In Eq.(\ref{eq:sigcorr}), the term $\sigma^{rad}$ includes the radiative 
corrections to the cross sections, calculated from the prescription of 
Bardin and Dokochueva \cite{Bar86}. The term $\sigma^{c}$ represents a 
correction for the finite charm quark mass since the data, particularly at 
low $Q^2$, is taken in a region close to charm quark threshold. 
The term $\sigma^{W}$ is a correction for the finite $W$ mass. The 
remaining correction, $\sigma^{iso}$, was an attempt to account for the neutron 
asymmetry in iron. We will review this correction in some detail.     

On average for the CCFR target, the neutron excess is given by 
$\nuf = (N-Z)/A = 0.0567$ \cite{Sel97a}.  
The CCFR group made an ``isoscalar correction'' to the 
neutrino and antineutrino cross sections. They calculated the quark and 
antiquark neutrino momentum densities on iron, via the $F_2$ structure 
functions per nucleon for $\nu$ and $\ovnu$ on a non-isoscalar target, 
\bea
&\,&xq^{\nu A}(x) = 2x\bigl[ \ZoA (d^p(x) + s^p(x)) \nonumber \\ &+& 
  \NoA (d^n(x) + s^n(x))\bigr] \nonumber \\ 
  &=& x(1-\nuf)d(x) + x(1+\nuf)u(x) + 2xs(x) ; 
  \nonumber \\ &\,&x\qbar^{\nu A}(x) = x(1-\nuf)\ubar(x) + x(1+\nuf)\dbar(x) ;
  \nonumber \\ &\,&xq^{\ovnu A}(x) = x(1-\nuf)u(x) + x(1+\nuf)d(x) ; 
  \nonumber \\ &\,&x\qbar^{\ovnu A}(x) = x(1-\nuf)\dbar(x) + x(1+\nuf)\ubar(x) 
  \nonumber \\ &\,& + 2x\sbar(x) \ .  
\label{eq:isoscalar}
\eea
The cross sections were then re-normalized by the ``isoscalar correction 
factor,'' 
\be
corr^{iso} \equiv \frac{\sigma({\rm isoscalar~target})}{\sigma({\rm 
 Fe~target})} = \frac{\sigma(\nuf = 0)}{\sigma(\nuf = 0.0567)}
\label{eq:isoscaldef}
\ee

The isoscalar correction is different for neutrinos and for antineutrinos. 
Note that this process is circular -- the isoscalar correction applied 
to the cross sections requires knowledge of the parton distributions, which 
are themselves extracted from the cross sections. Thus the process was 
applied iteratively. An isoscalar correction was applied to the cross 
sections from Eq.~(\ref{eq:isoscaldef}) using the parton distributions 
from Eq.~(\ref{eq:isoscalar}). The structure functions were then determined 
by inserting the renormalized cross sections into Eq.~(\ref{eq:Xsumdiff}). 
From the structure functions one can extract  
new parton distributions, from which a new isoscalar correction 
factor could be determined. The process was then iterated until the 
difference in the extracted structure functions became sufficiently small.   

The isoscalar correction factor applied by the CCFR collaboration 
should account for most of the neutron asymmetry corrections. 
However, after this correction has been applied it is then difficult 
to isolate the remaining contribution from the $\Delta F_2^{W A}$ term in 
Eq.~(\ref{eq:Xsumdiff}). Certainly there is a term present in the 
coupled equations, which has not been accounted for by the CCFR group. 
It should be straightforward to include this term in any re-analysis of 
neutrino cross section data. One could add this as a perturbation and 
could obtain decent estimates of the strange quark and CSV contributions 
as outlined in Sect.~\ref{Sec:Extract}. 

The sign and magnitude of the strange and CSV contributions should be  
similar to our estimate of these terms for an isoscalar target. There may also 
be small additional contributions from neutron asymmetry to $xF_3$, which have 
not been accounted for by the isoscalar correction made by CCFR. We have 
not made further corrections for any nuclear modification of the parton 
distributions in iron. Several groups have estimated the magnitude 
of nuclear effects on parton distribution functions \cite{Kum02,Hir04,Hir05,Kul06,Kul07,Kul07b,Cloet:2009qs}. 

\section{Conclusions\label{Sec:Concl}} 

The Gross-Llewellyn Smith sum rule is obtained from the first moment of 
the structure function $xF_3$ from neutrino charged-current deep inelastic 
scattering. In principle these structure functions can be obtained by 
comparing sums and differences of neutrino and antineutrino DIS cross 
sections on an isoscalar nucleus. Previous analyses of the GLS sum rule 
have neglected potential contributions from strange quark asymmetries and 
from partonic charge symmetry violation. At the time, such contributions 
were largely unknown and could be assumed to be negligibly small. 

However, recently one has more quantitative results for strange quark 
asymmetries from several groups \cite{Mason:2007zz,Martin:2009iq,Lai:2007dq,Ball:2009mk,Alekhin:2008mb}. All of these analyses rely on measurements of 
opposite-sign dimuon production in neutrino and antineutrino reactions on iron 
from the CCFR and NuTeV collaborations \cite{Baz95,Gon01}. All of these 
analyses obtained a positive value for $\langle xs^-(x)\rangle$; with the 
exception of tne NNPDF analysis \cite{Ball:2009mk} (which used a neural 
network approach, was relatively insensitive to the $s$ quark distribution 
and obtained very large error bars), these analyses found a crossover point 
for the $s^-(x)$ PDFs that occurred at an extremely small value $x_0 \approx 
0.01$. We chose as an example the strange quark analysis by the NuTeV group, 
Mason \EA~\cite{Mason:2007zz}, but the correction to the GLS sum rule arising 
from the strange quark asymmetry should be quite similar if one used 
instead the CTEQ \cite{Lai:2007dq}, MSTW \cite{Martin:2009iq} or Alekhin 
\cite{Alekhin:2008mb} analyses.  

Furthermore, one now has reasonable estimates for 
contributions from valence quark CSV. First, there are now phenomenological 
analyses of parton distributions that include partonic CSV \cite{MRST03}. 
Second, there have been calculations of partonic CSV arising from the 
different electromagnetic coupling of photons to up and down quarks 
\cite{MRST05,Glu05}. Finally, there are quark model calculations of 
partonic CSV \cite{Londergan:2009kj}. We used these to estimate the 
partonic CSV contribution to the Gross-Llewellyn Smith sum rule. 
Finally, we estimated the contribution to the GLS sum rule from the 
fact that iron is a non-isoscalar target. 

The correct procedure would be to incorporate these effects into the 
initial analysis of the neutrino cross sections. In particular one should 
take into account the effect of the term $\Delta F_2^{W A}(x)$ in 
Eq.~(\ref{eq:FApdf}). Since this term has not been included in previous 
analyses of neutrino cross sections we can only estimate its effect on 
the Gross-Llewellyn Smith sum rule. This has been carried out in this 
paper. To summarize our conclusions: first, the contributions from strange 
quarks, parton CSV and non-isoscalar effects all appear to have the same sign 
and hence to add coherently; second, we estimate that these effects 
should contribute an amount on the order of one to two standard deviations in 
the GLS sum rule; third, we find that inclusion of all of these contributions 
should bring the theoretical and experimental 
determinations of the GLS sum rule in agreement within $1\sigma$. 
In Sect.~\ref{Sec:GLSnuc} we noted that the cross section corrections 
adopted by the CCFR group make it difficult to provide quantitative 
estimates of the effects of strange quarks and partonic charge symmetry 
violation to the Gross-Llewellyn Smith sum rule. Nevertheless, if these 
corrections were to be integrated into a re-analysis of the neutrino cross 
sections, one should be able to obtain an accurate quantitative assessment 
of the contributions from these quantities. 

\acknowledgments
One author (AWT) acknowledges a useful discussion with S. Forte. One of the 
authors (JTL) was supported in part by the National Science 
Foundation under grant NSF PHY0854805. This work was also supported 
by the Australian Research Council through an Australian Laureate Fellowship
(AWT) and by the University of Adelaide.

\end{document}